\documentclass[twocolumn,showpacs,preprintnumbers,amsmath,amssymb,aps,showkeys]{revtex4}

\usepackage[dvips]{graphicx}
\usepackage{dcolumn}
\usepackage{bm}

\begin{document}



\title{Enhancing the yield of high-order harmonics with an array of gas jets}

\author{Angela Pirri$^{1}$, Chiara Corsi$^{1}$ and Marco Bellini$^{1,2}$ }
\affiliation{$^{1}$LENS, Via Nello Carrara 1, 50019 Sesto Fiorentino, Florence, Italy; $^{2}$Istituto
Nazionale di Ottica Applicata, Largo  E. Fermi 6, I-50125 Firenze, Italy}

\date{\today}

\begin{abstract}

We report the experimental observation of an enhancement in the yield of high-order harmonics using an
array of gas jets as the source medium. By comparing the experimental outcome for jet arrays of
different spacings with the predicted harmonic intensity in the case of slit sources of equivalent
lengths, we clearly show how the periodic modulation of the gas density can improve the harmonic yield.
This behavior may attributed to a quasi-phase-matching effect which increases the length of coherent
harmonic build-up during propagation by partially counteracting the dephasing induced by free electrons.

\end{abstract}


\maketitle

Since its discovery in 1987 \cite{bib1}, the phenomenon of high-order harmonic generation (HHG) has
become one of the most interesting topics in the field of highly nonlinear processes. Apart from its
fundamental physics interest, HHG is now one of the most promising ways to obtain tunable, short-pulse,
narrow-band radiation in the vacuum and extreme ultraviolet (VUV and XUV), and in the soft X-ray
regions, where other coherent sources are scarcely available. However, the possibility of using
high-order harmonics as a table-top VUV-XUV coherent source for applications is strongly connected to
the optimization of the brightness of the source over the broadest spectral range.

As in many other nonlinear processes, conversion efficiency in HHG depends on the interplay between the
single-atom response \cite{bib2} and the macroscopic response during propagation in the medium
\cite{bib3}. In particular, a constructive interference of the harmonic field emitted from different
locations of the source length can only be obtained over the so-called coherence length $L_c$. For media
longer than $L_c$, destructive interference soon depletes the generated field and dramatically limits
the conversion efficiency. By an appropriate choice of the interaction parameters it is often possible
to make the coherence length longer than the medium, thus reaching the so-called phase-matching
conditions. However, when phase-matching is not achievable, one can still artificially beat the
coherence length limits by properly modulating the interaction parameters along the field propagation
direction. Examples of such quasi-phase-matching (QPM) techniques now abound for low-order nonlinear
phenomena in structured crystals, and a few examples have been recently demonstrated in HHG. Many
different phenomena contribute to limit the coherence length in HHG. Here, the atomic dispersion and the
geometric Guoy phase \cite{bib4,bib5}, are always accompanied by the dispersion connected to free
electrons in the partially ionized medium \cite{bib6, bib6a}, and by the characteristic atomic dipole
phase \cite{bib7,bib8,bib9,bib10,bib11}. Depending on the gas type and density, and on the level of
ionization, the coherence length in HHG is usually so short as to pose a serious limit to significant
conversion efficiencies. Differently from QPM in periodically-poled nonlinear crystals \cite{bib12},
where the sign of the nonlinear coefficient is periodically flipped so as to always add constructively
interfering contributes to the generated field, QPM in HHG has so far been demonstrated by periodically
switching-off harmonic generation during the out-of-phase intervals. Either a modulation of the pump
laser intensity in a modulated-diameter, hollow-core, gas-filled waveguide \cite{bib14,bib15,bib16}, or
the use of a counter-propagating train of pulses \cite{bib17,bib17a} were used at this purpose.
Recently, theoretical studies have predicted the possibility of achieving QPM by periodically modulating
the density of the gas medium \cite{bib13,bib18}, and a proof-of-concept work has shown the coherent
buildup of the harmonic field in two separated gas sources \cite{bib19}.

Here we introduce a simple scheme for the generation of high-order harmonics from a whole array of
arbitrarily-spaced gas sources. We compare the experimental results to those one may expect from a
single-slit source of equivalent length and show that, for particular spacings and harmonic orders, a
clear enhancement of the yield is observed that may be attributed to at least a partial QPM effect.
\begin{figure}[h]
\includegraphics [width=8.5cm]{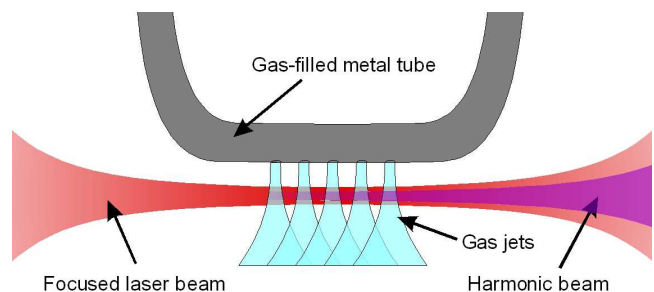}
\caption{Scheme of the interaction region for the generation of high-order harmonics. An array of
laser-drilled holes in a gas-filled metal tube produces a periodic modulation of the medium density
encountered by the focused pump laser pulses.} \label{holesc}
\end{figure}
Figure 1 shows a schematic of the experimental setup. The gas jets were obtained by laser drilling a
linear array of equally-spaced holes, each with a diameter of about 30 $\mu m$, over a total distance of
about 2 mm in a metal tube. The distance between the holes depends on their number: we used tubes with
three, five, ten and twenty holes, corresponding to separations of about 600, 400, 200 and 100 $\mu m$,
respectively. The 30-fs pulses from an amplified Ti:Sapphire laser were focused by a 50-cm focal-length
lens directly below the hole array while the tube was filled with xenon at low pressure. The peak
intensity in the focal spot, with a beam waist of 30 $\mu m$ and a corresponding confocal parameter of 5
mm (chosen to be longer than the global source length), was about $5.0 \times 10^{14} W/cm^2 $, well
above the saturation intensity of xenon. Intense plasma light emission was clearly observed from the
interaction zones under the tube holes (see Fig.2). The gas backing pressure in the tube was regulated
in order to maintain a constant density (estimated to be about 6.6 $\times 10^{17}atoms/cm^3$ and chosen
according to the maximum pressure allowed by the pumping system under continuous flow conditions) in the
interaction region of each gas jet for the different arrays. A single hole configuration operated at the
same gas density was used as a reference. The hole array was aligned to the laser propagation direction
and the tube was mounted on a translation stage allowing for precise positioning with respect to the
laser focus. The spectrum of the emitted harmonics was dispersed by a Pt-Ir normal incidence spherical
grating (600 lines/mm) and detected by a phosphor screen (converting the XUV-photons into visible light)
and a photomultiplier.

\begin{figure}[h]
\includegraphics [width=8.5cm]{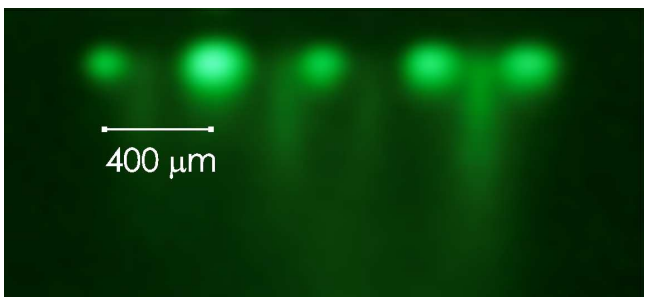}
\caption{Picture of the harmonic generation region in the 5-hole gas jet configuration; the plasma light
generated by intense laser ionization is evident in correspondence with the gas jets, separated by about
400 $\mu m$.} \label{holes}
\end{figure}
In order to clearly bring to evidence the effects of the gas jet array and to eliminate all systematic
contributions to the detected number of XUV photons (like the different reflectivity of the grating or
efficiency of the phosphor screen at different wavelengths), all the harmonic yields obtained in the
multiple-source configurations were normalized to the respective single-source ones. What we obtain for
each harmonic order is thus the effective enhancement factor due to the presence of the source array.
This enhancement factor can be simply compared to the theoretical expectation for a single slit source
emitting gas at the same density and with a length equivalent to the sum of the individual sources in
the array. Two limit situations are considered in this case: in the first one, the dispersion effects
due to the presence of free electrons are properly accounted for, limiting the coherence length of the
nonlinear interaction; in the second one, an infinite coherence length, corresponding to an unrealistic
perfect phase-matching, is assumed. The expected dependence of the harmonic yield as a function of the
interaction length $L$, the absorption length $L_a$, and coherence length $L_c$ is estimated according
to \cite{bib20}:
\begin{equation}
n_p(L,L_a,L_c) \propto \frac{L_a^2}{1+4 \pi^2 (\frac{L_a}{L_c})^2} [1+e^{-\frac{L}{L_a}}-2 \cos(\pi
\frac{L}{L_c}) e^{-(\frac {L}{2 L_a})}]
\end{equation}
and the corresponding enhancement factor for each harmonic order $q$ is simply given by:
\begin{equation}
E_q(n,L_{a_q},L_{c_q}) = \frac{n_p(n \cdot L,L_{a_q},L_{c_q})}{n_p(L,L_{a_q},L_{c_q})}
\end{equation}
where $n$ is the number of elementary sources, whereas $L_{a_q}$ and $L_{c_q}$ are the absorption and
coherence lengths calculated for each harmonic order by taking into account the source gas density and
always considering a full ionization of the medium \cite{bib21}. Since we are working in a relatively
loose focus configuration, ionization-related dispersion is the only factor taken into account for the
estimation of the coherence length. The medium length $L$ for a single gas jet source has been fixed to
45 $\mu m$. Note that the single-source length $L$ and the gas density are not independent and only
their product matters in the above estimations: increasing the source length has the same effect as
increasing the gas pressure by the same factor.
\begin{figure}[ht]
\includegraphics [width=8cm]{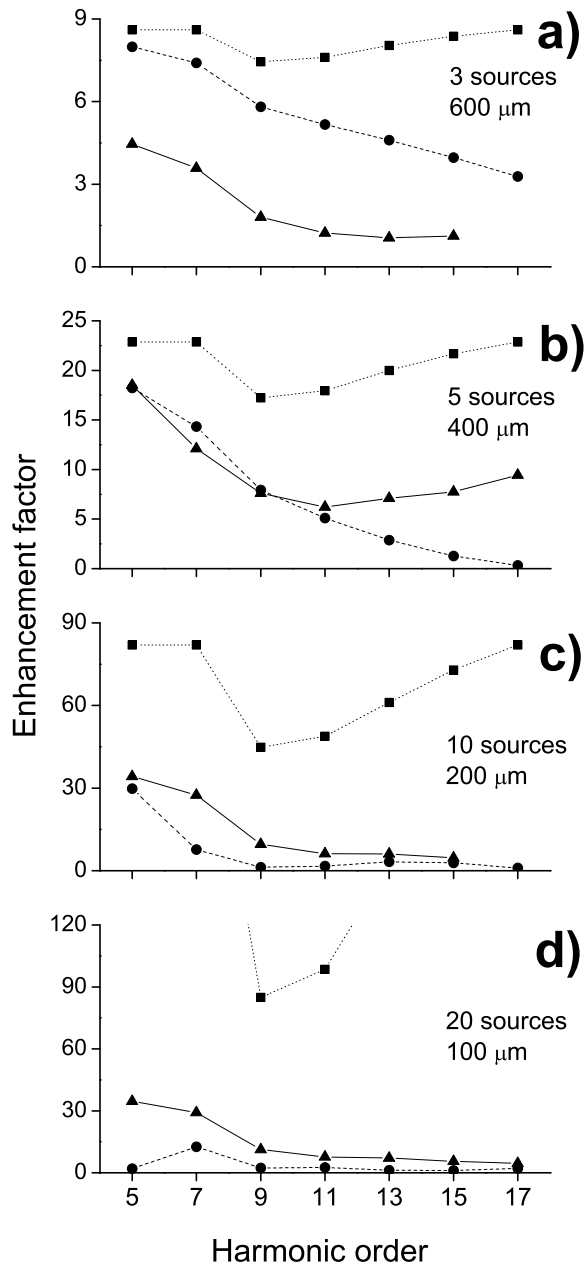}
\caption[] {Harmonic yield enhancement factors for different gas-jet source arrays normalized to the
single source configuration. Solid curves: experimental data; dotted curves: calculated data for a
single slit source of equivalent length in the case of perfect phase-matching; dashed curves: calculated
data for a single slit source including phase mismatch effects.}
\end{figure}

Simple considerations can be made before looking at the experimental data. First of all, if both
absorption and phase mismatch were absent, then the enhancement would be independent of the harmonic
order and proportional to the square of the medium length thanks to the constructive interference of all
emitters. In other words, one should expect an enhancement factor just equal to the square of the number
$n$ of sources in the array. Rather interestingly, such behavior is almost respected for our
lowest-order harmonics. In the case of the fifth and seventh harmonics, which are just below the
single-photon ionization threshold of xenon, a low absorption and a relatively long coherence length
contribute to reach high experimental enhancement factors, up to about 18, or about $70\%$ of the ideal
case for the fifth harmonic in the 5-hole array (see Fig.3b). Enhancement factors up to 35 were observed
for the fifth harmonic also in the 10 and 20-hole configurations (Figs.3c and 3d), but in these cases
the ideal absorption- and mismatch-free enhancements would have been much higher.

When absorption processes are turned on, the maximum enhancement factor decreases from the ideal value
of $n^2$, and starts to strongly depend on the harmonic order. The dotted curves in Fig.3 show this
expected behavior, with the pronounced dip due to absorption for harmonics just above the ionization
threshold. Finally, when phase mismatch is considered, a strong suppression of higher-order harmonics
and wild oscillations for lower orders are expected (dashed curves in Fig.3).

Experimental results are presented in Fig.3 (solid curves) for the four array configurations together
with the calculated curves for the long slit case, with and without dispersion effects. In most of the
cases for the 10 and 20-hole arrays (corresponding to source spacings of 200 and 100 $\mu m$,
respectively), the experimental points stay well above the curves corresponding to the non-phase-matched
situation. For the 5-hole array (400 $\mu m$ spacing), the expected mismatched behavior is exactly
followed by the lower orders, whereas a clear departure from this trend is obtained above the $11^{th}$
harmonic. In particular, a 30-fold increase of the enhancement factor is obtained for the $17^{th}$
harmonic, which should have been severely suppressed by a limited coherence length in the mismatched
situation. Although the enhancement levels expected for a completely phase-matched situation are never
reached, the observed behaviors suggest that at least a partial quasi-phase-matching effect is present.

The experimental points for the 3-hole array have a similar trend as the predicted mismatched situation,
but are always lower by a factor of about 2. Besides indicating that no QPM effect is probably present
in this case, the low enhancement factor may indicate some imperfect alignment of the rather distant
(600 $\mu m$ spacing) holes along the laser direction, or some strong effect of beam defocusing in the
heavily ionized medium~\cite{bib6a}. Although a similar phenomenon should be present also in the other
array configurations, the closer spacing of the sources and a partial QPM might be able to counteract
it.

As a final check, we also repeated the above measurements while varying the relative position between
the gas source arrays and the laser focus in a 5 mm range around the optimum. No significant deviations
from the above results were observed in such conditions, showing that the (position-insensitive)
free-electron contribution to dispersion is the most important in these cases and can be effectively
counteracted independently of the position.

In conclusion, the use of a periodically-modulated gas source has produced a clear enhancement in the
yield of high-order harmonics. Such an enhancement, dependent on the harmonic order and on the period of
the gas-density modulation, exceeds that predicted for a single slit source of equivalent length where
phase-matching effects, mostly due to the presence of free electrons, are present. This strongly
suggests a quasi-phase-matching effect connected to the periodicity along the propagation axis. The
experimental scheme, using an array of gas jets operating in a continuous flow, is extremely simple and
can be easily extended to longer interaction lengths or closer source spacings in order to further
increase the harmonic yield or enhance shorter wavelengths. Additional analysis is in progress, but
these results already show that a selective enhancement of the source brightness can be achieved in
order to optimize specific applications.


\end{document}